\documentclass[12pt,preprint]{aastex}

\shorttitle{Temporal Self-Similar MHD Jets}
\shortauthors{Tsui}
\begin{document}
\title{Time-Dependent Magnetohydrodynamic Self-Similar
\\
 Extragalactic Jets}
\author{K.H. Tsui and A. Serbeto}
\affil{Instituto de F\'{i}sica - Universidade Federal Fluminense
\\Campus da Praia Vermelha, Av. General Milton Tavares de Souza s/n
\\Gragoat\'{a}, 24.210-346, Niter\'{o}i, Rio de Janeiro, Brasil.}
\email{tsui$@$if.uff.br}
\date{}
\pagestyle{myheadings}
\baselineskip 18pt
	 
\begin{abstract}

Extragalactic jets are visualized as dynamic erruptive events modelled
 by time-dependent magnetohydrodynamic (MHD) equations.
 The jet structure comes through the temporally self-similar solutions
 in two-dimensional axisymmetric spherical geometry.
 The two-dimensional magnetic field is solved in the finite plasma
 pressure regime, or finite $\beta$ regime, and it is described by an
 equation where plasma pressure plays the role of an eigenvalue.
 This allows a structure of magnetic lobes in space, among which the
 polar axis lobe is strongly peaked in intensity and collimated in
 angular spread comparing to the others. For this reason, the polar
 lobe overwhelmes the other lobes, and a jet structure arises in
 the polar direction naturally. Furthermore, within each magnetic
 lobe in space, there are small secondary regions with closed
 two-dimensional field lines embedded along this primary lobe.
 In these embedded magnetic toroids, plasma pressure
 and mass density are much higher accordingly. These are termed as
 secondary plasmoids. The magnetic field lines in these secondary
 plasmoids circle in alternating sequence such that adjacent plasmoids
 have opposite field lines. In particular, along the polar primary lobe,
 such periodic plasmoid structure happens to be compatible with radio
 observations where islands of high radio intensities are mapped.

\end{abstract}

\vspace{2.0cm}

\maketitle
\newpage
\section{Introduction}

Collimated jets with high terminal velocities appear to be universal
 phenomena in astrophysics. They are often associated
 with young stellar objects, compact galactic objects, and active
 galactic nuclei [Livio 1997]. These jets are always accompanied by
 accretion disks on the equatorial plane. The magnetosphere of the
 accretion disk contains a plasma that follows the accretion of the
 materials in the disk. The plasma density and pressure get higher as
 the accretion approaches to the center which drive the magnetic field.
 The dynamics of this system was first described in magnetohydrodynamic
 (MHD) model by Blandford and Payne [1982]. In this landmark paper,
 jets are considered as a spatial structure in stationary state.  
 Ideal MHD equations in cylindrical coordinates $(r,\phi,z)$ are solved
 for time independent solutions. The central mass $M$ is replaced by a
 linear mass along the cylindrical axis. Self-similar solutions in
 space with a scale invariance $z/r$ are sought. Such self-similar
 solutions are compatible to a Keplerian disk plasma rotation velocity
 field superimposed by an Alfv\'{e}nic plasma velocity. The complete
 MHD instability spectrum with such Keplerian profile are analysed by
 Keppens, Casse, and Goedbloed [2002]. This model predicts that
 the magnetospheric disk plasma would be ejected towards the polar
 direction magnetocentrifugally should the magnetic field lines be at
 an angle less than $\pi/3$ or more that $2\pi/3$ on the $rz$ plane
 with respect to the outward radius of the disk. Collimating action
 of this plasma outflow would be provided by the hoop force of the
 azimuthal magnetic field and its associated parallel current in a
 force-free configuration. The interaction of the magnetic field with
 the plasma disk generates a MHD Poynting flux [Ferreira and Pelletier
 1995] that can be converted into kinetic energy of the jet plasma
 [Zanni et. al. 2004]. Variants of jet formation model are proposed
 by Contopoulos and Lovelace [1994], Contopoulos [1995], and Cao and
 Spruit [1994]. This accretion-ejection model provides the basic
 framework of current investigations of accretion disk and jets as
 is reviewed by Balbus and Hawley [1998]. Dissipative MHD effects
 are examined by Casse and Ferreira [2000 a,b] and Casse [2004], and
 relativistic jets are analyzed by Vlahakis [2004]. These stationary
 state analytic studies are often complemented by numerical works
 in cylindrical geometry to simulate time evolutions of the jets
 [Ustyugova et. al. 1995, Ouyed and Pudritz 1997, Krasnopolski et. al.
 1999], and the disk-jet system [Matsumoto et. al. 1996,
 Kato et. al. 2002].

Here, instead of a stationary model, we take the view that jets are
 a time dependent spatial structure. What we see is only a snapshot
 of their state at this particular moment. Due to their galactic
 dimensions, the time scale of these structures is believed to be
 extraordinarily large which gives the impression of a stationary
 structure. This implies that jets are results of an erruption
 originating from the galactic nucleus. They could be dissipated in
 time before another erruption takes place due to pressure built-up
 from accretion. Or one erruption could be superimposed on an earlier
 event. To model the jet system, we will do a self-similar analysis
 on the full time-dependent ideal MHD equations in spherical
 coordinates $(r,\theta,\phi)$ with a mass $M$ at the nucleus.
 In particular, we consider axisymmetric solutions. This type of
 self-similar solutions were pioneered by Low [1982a,b, 1984a,b]
 for astrophysical applications and solar corona mass ejections
 with pure radial plasma velocity flow. Variants of these solutions
 include cases where the plasma domain lies outside the mass $M$
 such as interplanetary magnetic ropes in one-dimensional
 [Osherovich, Farrugia, and Burlaga 1993, 1995] and in two-dimensional
 [Tsui and Tavares 2005] cylindrical geometry, interplanetary magnetic
 clouds [Tsui 2006a], and also atmospheric ball lightnings [Tsui 2006b]
 in spherical geometry. In these descriptions where $M$ is outside
 the spherical domain of interest, the magnetic field is axisymmetric
 force-free and contains regions of closed field lines, while the
 plasma is spherically symmetric decoupled from the magnetic field.
 
For the present case of extragalactic jets described by the mechanism
 of accretion-ejection, we will follow the self-similar solutions
 of Low with the polytropic index $\gamma=4/3$, but with particular
 emphasis on the finite plasma pressure. This current approach
 differs from the Keplerian disk plasma in that the radial flow
 is not tied to the Alfv\'{e}n velocity as a priori.
 In this dynamic model, we consider jets as a manifestation of mass
 ejection on a galactic scale. The plasma pressure, in this self-similar
 MHD model, proves to have an important role in collimating the magnetic
 fields and the jet plasmas. The time evolution function gives a
 dynamic description of the high radial flow especially in the jets.
 The self-similar solutions, that converge at the center and at
 infinity, give small regions of closed axisymmetric two-dimensional
 magnetic field lines where plasma density and pressure are much higher.
 These regions along the jets could correspond to the high intensity
 islands in radio frequency maps.
 
\newpage
\section{Self-Similar MHD}

Following the accretion-ejection classical model, we also use the
 MHD equations to describe the plasma. Nevertheless, we retain the
 time dependence to write
\\
$${\partial\rho\over\partial t}+\nabla\cdot(\rho\vec v)\,
 =\,0\,\,\,,\eqno(1)$$

$$\rho\{{\partial\vec v\over\partial t}
 +(\vec v\cdot\nabla)\vec v\}\,
 =\,\vec J\times\vec B-\nabla p-\rho{GM\over r^3} \vec r
 \,\,\,,\eqno(2)$$

$${\partial\vec B\over\partial t}\,
 =\,-\nabla\times\vec E\,
 =\,\nabla\times(\vec v\times\vec B)\,\,\,,\eqno(3)$$

$$\nabla\times\vec B\,=\,\mu\vec J\,\,\,,\eqno(4)$$

$$\nabla\cdot\vec B\,=\,0\,\,\,,\eqno(5)$$

$${\partial\over\partial t}({p\over\rho^{\gamma}})
 +(\vec v\cdot \nabla)({p\over\rho^{\gamma}})\,
 =\,0\,\,\,.\eqno(6)$$
\\
Here, $\rho$ is the mass density, $\vec v$ is the bulk velocity,
 $\vec J$ is the current density, $\vec B$ is the magnetic field,
 $p$ is the plasma pressure, $\mu$ is the free space permeability,
 and $\gamma$ is the polytropic index.
 For the bulk velocity, it is consisted of a radial and an azimuthal
 component to model the plasma outflow and the disk rotation.
 To be compatible to the physical situation, the meridian velocity
 is taken to be null. For the radial component $v\hat r$, we seek
 self-similar solutions where the time evolution is described by
 the dimensionless evolution function $y(t)$. The radial profile
 is time invariant in terms of the radial label $\eta=r(t)/y(t)$,
 which has the dimension of $r$, such that $\eta$ is independent
 of time, and corresponds to the Lagrangian radial position attached
 to a fixed fluid element. The label $\eta$ corresponds to the
 radial positions of the initial self-similar configuration that
 expands in time with radial Eulerian positions $r(t)=\eta y(t)$.
 The velocity can then be written as
\\
$$\vec v\,=\,\{\eta{dy\over dt}, 0, v_{\phi}\}\,\,\,.\eqno(7)$$
\\
We consider a two-dimensional case with azimuthal symmetry in
 $\phi$. In this case, the magnetic field, through the vector
 potential $\vec A$, can be expressed in terms of two scalar
 functions $P$ and $Q$
\\
$$\vec B\,=\,{1\over r\sin\theta}
 \{+{1\over r}{\partial\over\partial\theta}(rA_{\phi}\sin\theta),
 -{\partial\over\partial r}(rA_{\phi}\sin\theta),
 +\sin\theta[{\partial\over\partial r}(rA_{\theta})
 -{\partial\over\partial\theta}(A_{r})]\}\,$$

$$=\,{1\over r\sin\theta}
 \{+{1\over r}{\partial\over\partial\theta}(P),
 -{\partial\over\partial r}(P),
 +Q\}\,
 =\,\nabla P\times\nabla\phi+Q\nabla\phi\,\,\,.\eqno(8)$$
\\
We remark that the velocity field in cylindrical stationary state
 accretion-ejection model is a three-component field. This is
 necessary because the jets are generated by magnetocentrifugal
 motion of the planar disk plasma to the axial direction.
 Here, in this spherical dynamic model, the velocity field is a
 two-component field with $v_{\theta}=0$, because the jets are
 generated by first pulling the disk plasma to the center and
 then redirecting it to space radially through erruptions.

By self-similar solutions in time, we mean a special
 class of time-dependent solutions where the time and space parts
 of the physical quantities are in a separable form. The time part
 is described by the evolution function $y(t)$, and the space part
 will be solved self-consistently by separation of variables.
 The concept of self-similar dynamics is closely related
 to self-organized states that often have minimum energy under
 given constraints. Having this in mind, we now transform the independent
 variables from $(r,\theta,t)$ to $(\eta,\theta,y)$, and proceed to
 determine the explicit dependence of $y$ in each one of the physical
 quantities with this radial velocity. First, making use of Eq.(7),
 Eq.(1) renders
\\
$${\partial\rho\over\partial t}
 +{1\over r^2}{\partial\over\partial r}(r^2v\rho)\,
 =\,({\partial\rho\over\partial t}
 +v{\partial\rho\over\partial r})
 +\rho({\partial v\over\partial r}+{2v\over r})\,
 =\,0\,\,\,.\eqno(9a)$$
\\
Considering the second equality, the first bracket is the convective
 time derivative in Euler fluid coordinates, and this amounts to
 the time derivative in Lagrangian fluid coordinates. We, therefore,
 have
\\
$${\partial\rho\over\partial y}{dy\over dt}
 +\rho({\partial v\over\partial r}+{2v\over r})\,
 =\,{\partial\rho\over\partial y}{dy\over dt}
 +{3\rho\over y}{dy\over dt}\,
 =\,({\partial\rho\over\partial y}+{3\rho\over y})
 {dy\over dt}\,=\,0\,\,\,.$$
\\
Solving this equation for the $y$ dependence gives
\\
$$\rho(\vec r,t)\,
 =\,{1\over y^3}\bar\rho(\eta,\theta)\,\,\,.\eqno(9b)$$
\\
As for Eq.(6), with $F=(p/\rho^{\gamma})$ it follows that
\\
$${\partial F\over\partial t}+v{\partial F\over\partial r}\,
 =\,{\partial F\over\partial y}{dy\over dt}\,
 =\,0\,\,\,,\eqno(10a)$$

$$({p\over\rho^{\gamma}})\,=\,F(\vec r,t)\,
 =\,{1\over y^0}\bar F(\eta,\theta)\,\,\,.\eqno(10b)$$
\\
Using the representation of Eq.(8), Eq.(3) is represented by
 the following two equations
\\
$${\partial P\over\partial t}+v{\partial P\over\partial r}\,
 =\,{\partial P\over\partial y}{dy\over dt}\,
 =\,0\,\,\,,\eqno(11a)$$

$${\partial Q\over\partial t}+{\partial\over\partial r}(vQ)\,
 =\,{\partial\over\partial r}
 (v_{\phi}{1\over r}{\partial P\over\partial\theta})
 -\sin\theta{\partial\over\partial\theta}
 (v_{\phi}{1\over r\sin\theta}{\partial P\over\partial r})
 \,\,\,.\eqno(12)$$
\\
The first equation gives the function $P$ as
\\
$$P(\vec r,t)\,
 =\,{1\over y^0}\bar P(\eta,\theta)\,\,\,.\eqno(11b)$$
\\
As for the function $Q$, the right side of the second equation
 vanishes for rigid rotation to give
\\
$$v_{\phi}\,=\,\omega_{0}r\sin\theta\,\,\,,\eqno(13)$$

$${\partial Q\over\partial t}+{\partial\over\partial r}(vQ)\,
 =\,{\partial Q\over\partial y}{dy\over dt}
 +{Q\over y}{dy\over dt}\,
 =\,({\partial Q\over\partial y}+{Q\over y})
 {dy\over dt}\,
 =\,0\,\,\,,\eqno(14a)$$

$$Q(\vec r,t)\,
 =\,{1\over y^1}\bar Q(\eta,\theta)\,\,\,.\eqno(14b)$$
\\
We note that $v_{\phi}$ amounts to a constant plasma rotation,
 which has to be distinguished from the Keplerian disk rotation
 of the solid material.
 Comparing to the erruptive time scale of our dynamic approach of
 jets, this rotation rate is negligible and we take $\omega_{0}=0$.
 As for the magnetic field components, they are given by 
\\
$$B_{r}\,=\,{1\over y^2}\bar B_{r}(\eta,\theta)\,
 =\,+{1\over y^2}{1\over\eta\sin\theta} mbbbbbbb
 {1\over\eta}{\partial\bar P\over\partial\theta}
 \,\,\,,$$ 

$$B_{\theta}\,=\,{1\over y^2}\bar B_{\theta}(\eta,\theta)\,
 =\,-{1\over y^2}{1\over\eta\sin\theta}
 {\partial\bar P\over\partial\eta}
 \,\,\,,$$ 

$$B_{\phi}\,=\,{1\over y^2}\bar B_{\phi}(\eta,\theta)\,
 =\,+{1\over y^2}{1\over\eta\sin\theta}\bar Q
 \,\,\,.$$ 
\

\newpage
\section{Self-Organization}

Before proceeding with the analysis, let us recall on the
 fundamentals of self-organiation, in particular on MHD systems.
 We note that MHD equations, like Navier-Stokes equations, have
 quadratic invariants in the absence of dissipations.
 In MHD systems, there are three of them. They are the
 total energy (plasma and magnetic), magnetic helicity, and
 cross-helicity. Because of the existence of multiple invariants,
 the system tends to develope self-organized and self-similar states
 through dissipative processes regardless of the details of the
 initial conditions [Hasegawa 1985]. A simple example in fluid
 mechanics is the developement of shock waves from an initial
 explosion. Another example is that fluid vortex rings (solitons)
 in air are often formed in fast upward drafts of smokes. Should
 we consider a thin layer of oil heated from below, a grid of
 complex highly organized hexagonal convective cells would
 develope regardless of the details of the initial conditions.
 We note that, for self-organized states, the memories of the
 initial conditions are lost. In other words, we can not trace a
 self-organized state backwards in time to its initial conditions.
 They are lost in the dissipative processes that lead to organization.
 The underlying physical arguments for self-similar solutions are
 also discussed in detail by Low [1982a]. For these fundamental
 reasons, although complex self-similar solutions are only a subset
 of general time-dependent MHD solutions, where most of them are not
 self-similar, they are prone to develope in nature with simple
 initial configurations.

Numerically, starting from MHD equations with any fluctuations in a
 given initial configuration, self-organization could be reached
 since they are insentive to the details in the initial conditions.
 Analytically, the nature of self-similarity implies that the dependent
 variables $(\vec r,t)$ of a physical variable appear in separable form
 as we have done in Eqs.(9-14). As a consequence, the solutions will
 be obtained by the method of separation of variables. Naturally,
 this imposes severe restrictions of the physical system where such
 a procedure is feasible, such as the dimensionality and symmetry.
 Most of the self-similar solutions are established in axisymmetric
 systems whether cylindrical or spherical. Nevertheless, Gibson and
 Low [1998] have made a great leap in establishing a three-dimensional
 spherical solution. Under the framework of separation of variables,
 different kinds of solutions can be obtained for the same system
 depending on the choice of constants. Likewise in MHD systems,
 we can have different self-similar solutions to account for
 different phenomena depending on how we separate the constants.
 In the next section, we choose to solve the system with an oscillating
 radial solution because this solution gives magnetic toroids along the
 jet that match with observations. A monotonic decreasing or increasing
 radial solution in power form is also possible [Lynden-Bell and Boily 1994].
 Although this monotonic solution is not relevant for galactic jets,
 it could be useful for other natural phenomena such as in two-dimensional
 interplanetary magnetic ropes [Tsui and Tavares 2005] and some other
 astrophysical objects.
 
\newpage
\section{Low Model}

Comparing the equation of $F=(p/\rho^{\gamma})$ and the equation
 of $P$, we conclude that $F=F(P)$ is a functional of $P$ or
 $\bar F=\bar F(\bar P)$. We remark that usually we can not make
 the above statement just based on the similarity of the governing
 equations. It is only possible when we are under the framework
 of self-similarity. We can, therefore, write the $\theta$
 dependence in $\bar\rho$ and $\bar p$ in terms of $\bar P$
 to get
\\
$$\bar\rho\,=\,\bar\rho(\eta,\bar P)\,\,\,,\eqno(15a)$$

$$p\,
 =\,{1\over y^{3\gamma}}\bar F(\eta,\theta)
 \bar\rho^{\gamma}(\eta,\theta)\,
 =\,{1\over y^{3\gamma}}\bar F(\bar P)
 \bar\rho^{\gamma}(\eta,\bar P)\,
 =\,{1\over y^{3\gamma}}\bar p(\eta,\bar P)\,\,\,.\eqno(15b)$$
\\
Furthermore, the $\eta$ and $\bar P$ dependences should be in a
 separable form in both
\\
$$\bar\rho(\eta,\bar P)\,
 =\,\bar\rho_{1}(\eta)\bar\rho_{2}(\bar P)\,\,\,,$$

$$\bar p(\eta,\bar P)\,
 =\,\bar p_{1}(\eta)\bar p_{2}(\bar P)\,\,\,,$$
\\
so that, with an adequate $\gamma$, $\bar F$ could come out as a
 functional of $\bar P$ only. Making use of Eq.(4) to eliminate
 the current density in Eq.(2), we get the momentum equation
 which has three components. First, we examine the $\phi$ component
 which contains only the magnetic force
\\
$${\partial P\over\partial r}
 {\partial Q\over\partial\theta}
 -{\partial P\over\partial\theta}
 {\partial Q\over\partial r}\,
 =\,{1\over y^2}
 \{{\partial\bar P\over\partial\eta}
 {\partial\bar Q\over\partial\theta}
 -{\partial\bar P\over\partial\theta}
 {\partial\bar Q\over\partial\eta}\}\,
 =\,0\,\,\,,\eqno(16a)$$
 
$$\bar Q\,=\,\bar Q(\bar P)\,\,\,.\eqno(16b)$$ 
\\         
The vanishing of the magnetic force in the azimuthal direction
 described by Eq.(16a) implies the functional relationship
 given by Eq.(16b). As for the $\theta$ component, with the
 knowledge of $p=\bar p(\eta,\bar P)/y^{3\gamma}$, it reads
\\
$${\partial P\over\partial\theta}
 \{{\partial^2 P\over\partial r^2}
 +{1\over r^2}\sin\theta{\partial\over\partial\theta}
 ({1\over\sin\theta}{\partial P\over\partial\theta})
 +Q{\partial Q\over\partial\theta}
 /{\partial P\over\partial\theta}
 +\mu r^2\sin^2\theta
 {\partial p\over\partial\theta}
 /{\partial P\over\partial\theta}\}\,$$
 
$$=\,{1\over y^2}{\partial\bar P\over\partial\theta}
 \{{\partial^2 \bar P\over\partial\eta^2}
 +{1\over\eta^2}\sin\theta{\partial\over\partial\theta}
 ({1\over\sin\theta}{\partial\bar P\over\partial\theta})
 +\bar Q{\partial\bar Q\over\partial\bar P}
 +\mu y^{4-3\gamma}\eta^2\sin^2\theta
 {\partial\bar p\over\partial\theta}
 /{\partial\bar P\over\partial\theta}\}\,
 =\,0\,\,\,.\eqno(17a)$$
\\
We remark that the first three terms of this equation represent
 the force-free field equation for $\vec J\times\vec B=0$
 [Aly 1984, Low 1986, Low and Lou 1990, Lynden-Bell and Boily 1994].
 The last term is the plasma pressure. This equation would be
 independent of the evolution function $y$ should we consider the
 $\gamma=4/3$ case pioneered by Low.

We note that a nonlinear equation like Eq.(17a) has to be solved
 subject to the boundary conditions of a given physical problem.
 For example, should the problem on hand has an infinite external
 domain and power form radial solutions are physically reasonable,
 then we use fractional powers of $\bar P$ for the functionals
 of $\bar Q$ and $\bar p$ [Lynden-Bell and Boily 1994]. In our
 present case, we are interested in decaying oscillating solutions
 in $\eta$. We, therefore, write
\\
$$\bar Q(\bar P)\,=\,a\bar P\,\,\,,\eqno(18a)$$
 
$$\bar p(\eta,\bar P)\,=\,\bar p_{1}(\eta)\bar p_{2}(\bar P)\,
 =\,(\eta^{-4})(b'^{2}\bar P^{2}+\bar C)\,\,\,,\eqno(18b)$$
\\
where $\bar C$ is a positive constant independent of coordinates
 and the functional $\bar P$. This is the simplest representation
 where the first equation gives a linear dependence of $\bar P$
 and the second equation reflects the positive definite nature
 of plasma pressure. This choice of constant $\bar C$ in plasma
 pressure profile differs from the spherically symmetric radial
 coordinate dependent additive term in Eqs.(21) and (22) of Low
 [1984b]. This difference of representation stems from the view
 that Low considers Eq.(17a) as an equation that solves for the
 plasma pressure under a given $\bar P$. For this reason,
 the homogeneous solution corresponds to the spherically symmetric
 gasodynamic solution. We regard Eq.(17a) as an equation that
 solves for $\bar P$ under a given plasma pressure that has
 a positive definite separable form as in Eq.(18b). The choice of
 Eq.(18a) gives $\partial\bar Q/\partial\bar P=a$, a constant.
 With $b^2=2\mu b'^2$, Eq.(17a) now reads
\\
$$\eta^2{\partial^2 \bar P\over\partial\eta^2}
 +\sin\theta{\partial\over\partial\theta}
 ({1\over\sin\theta}{\partial\bar P\over\partial\theta})
 +\eta^2 a^2\bar P
 +b^2\sin^2\theta\bar P\,=\,0\,\,\,.\eqno(17b)$$
\\
Writing $\bar P(\eta,\theta)=R(\eta)\Theta(\theta)$,
 $x=\cos\theta$, and with $n(n+1)$ as separation constant,
 Eq.(17b) becomes
\\
$$\eta^2{\partial^2 R\over\partial\eta^2}
 +[a^2\eta^2-n(n+1)]R\,=\,0\,\,\,,$$

$$(1-x^2){d^2\Theta(x)\over dx^2}
 +[n(n+1)+b^2(1-x^2)]\Theta(x)\,=\,0\,\,\,.$$
\\
The $R(\eta)$ equation can be solved readily to give
\\
$$R(\eta)\,=\,(a\eta)^{1/2}J_{n+1/2}(a\eta)\,\,\,.\eqno(19a)$$
\\
Such a spherical Bessel functional was used by Low [1984b] in
 Eq.(28) of his paper as one of the numerical examples in
 spherical two-dimensional self-similar MHD to model coronal
 mass ejections.
 
The $\Theta(x)$ equation with finite plasma pressure, $b^2\neq 0$,
 can be solved by power series
\\
$$\Theta(x)\,=\,\sum\,a_{m}x^{m}\,\,\,,\eqno(19b)$$

$$(m+2)(m+1)a_{m+2}=[m(m-1)-n(n+1)-b^2]a_{m}+b^2a_{m-2}\,\,\,,$$

$$6a_{3}=-[n(n+1)+b^2]a_{1}\,\,\,,$$

$$2a_{2}=-[n(n+1)+b^2]a_{0}\,\,\,,$$
\\
where the infinite sum in Eq.(19b) starts from $m=0$. There are
 two independent solutions. The first one has
 $a_{0}=\Theta(0)\neq 0$ and $a_{1}=0$, and the second one has
 $a_{0}=0$ and $a_{1}=d\Theta(0)/dx\neq 0$. They correspond to
 even and odd powers of the series respectively
\\
$$\Theta_{even}(x)\,=\,\sum\,a_{2m}x^{2m}\,\,\,,$$

$$\Theta_{odd}(x)\,=\,\sum\,a_{2m+1}x^{2m+1}\,\,\,.$$
\\
In the absence of plasma pressure with $b^2=0$, the above
 solution reduces to 
\\
$$\Theta(x)\,=\,(1-x^2){dP_{n}(x)\over dx}\,
 =\,-n(n+1)\int_{1}^{x}P_{n}(x)dx\,\,\,,\eqno(19b')$$
\\
where $P_{n}(x)$ is the Legendre polynomial. The even solution
 corresponds to $n$ odd, and the odd solution is
 otherwise. 
 
\newpage
\section{Self-Similar Magnetic Field}

With the solution $\bar P(\eta,\theta)=R(\eta)\Theta(\theta)$
 established by Eqs.(19), the magnetic field components are
\\
$$B_{r}\,=\,-{1\over y^2}{1\over\eta^2}R(\eta)
 {d\Theta(x)\over dx}\,\,\,,$$

$$B_{\theta}\,=\,-{1\over y^2}{1\over\eta}{dR(\eta)\over d\eta}
 {1\over (1-x^2)^{1/2}}\Theta(x)\,\,\,,$$

$$B_{\phi}\,=\,+{1\over y^2}{a\over\eta}R(\eta)
 {1\over (1-x^2)^{1/2}}\Theta(x)\,\,\,.$$
\\
We note that the self-similar evolution of the MHD plasma distorts
 the dipole-like magnetic field by generating an azimuthal
 component of the magnetic field with finite $a^2$.
 To grasp the magnetic structure given by Eqs.(19a) and (19b),
 we first examine the case with $b^2=0$. In this special case,
 the even and odd power series of Eq.(19b) will terminate at a
 finite number of terms when $m=n+1$ to give Eq.(19b') and
 $\Theta(x)=0$ at the poles $x=\pm 1$.

The self-similar radial structure $R(\eta)$ given by Eq.(19a)
 allows oscillations in $\eta$ if $a\eta$ is sufficiently
 large, which means that the azimuthal magnetic field is
 sufficiently large. The meridian structure $\Theta(x)$ given by
 Eq.(19b') also oscillates in $x$. Let us denote $\eta_{i}$ and
 $x_{j}$ as where $R(\eta)$ and $\Theta(x)$ vanish. We remark that
 $\eta_{i}$ are circles of constant $\eta$, and $x_{j}$ are spokes
 of constant $x=\cos\theta$. Consequently, $(\eta_{i},x_{j})$
 divide the $(\eta-x)$ plane in many smaller regions.
 On $\eta_{i}$, we have $B_{r}=0$ and $B_{\phi}=0$, with
 $B_{r}$ and $B_{\phi}$ changing signs across $\eta_{i}$.
 The only component that does not vanish completely on $\eta_{i}$
 is $B_{\theta}$. Referring to the complete expression of $B_{\theta}$
 above, this magnetic component is modulated by $\Theta(x)$ so that
 it changes sign on the circle of constant $\eta$ on crossing each
 spoke region of $x_{j}$.
 On $x_{j}$, we have $B_{\theta}=0$ and $B_{\phi}=0$, with
 $B_{\theta}$ and $B_{\phi}$ changing signs across $x_{j}$.
 The only component that does not vanish completely on $x_{j}$
 is $B_{r}$. Referring to the complete expression of $B_{r}$
 above, this magnetic component is modulated by $R(\eta)$ so that
 it changes sign on the spoke of constant $x$ on crossing each
 circular region of $\eta_{i}$.
 As a result, axisymmetric closed magnetic field lines are formed
 in these regions generating toroidal belt plasmoids circumscribing
 the z-axis of symmetry. We call these secondary plasmoids, and
 call the larger self-similar plasmoid embedding all the secondary
 plasmoids the primary plasmoid. Neighboring plasmoids have field
 lines circling in opposite sense. If one plasmoid has field lines
 in clockwise direction, the adjacent one has them in counter
 clockwise direction. The azimuthal components also rotate against
 each other.

In each region bounded by
 $(\eta_{i},\eta_{i+1})$ and $(x_{j},x_{j+1})$,
 the topological center defined by $dR(\eta)/d\eta=0$ and
 $d\Theta(x)/dx=0$ has $B_{r}=0$ and $B_{\theta}=0$. This is the
 magnetic axis of each toroid. The field lines about this center
 are given by
\\
$${B_{r}\over dr}\,=\,{B_{\theta}\over rd\theta}\,
 =\,{B_{\phi}\over r\sin\theta d\phi}\,\,\,.\eqno(20a)$$
\\
By axisymmetry, the magnetic field components are independent
 of $\phi$. For this reason, the third group of the above equation
 is decoupled from the first two groups. The $B_{\phi}$ field
 circles about the z-axis of symmetry without twisting. It is
 simply superimposed on the $B_{r}-B_{\theta}$ field lines.
 In terms of Fourier components $e^{im\phi}$, this means $m=0$.
 For the field lines on an $(r-\theta)$ plane, we consider the
 first equality between $B_{r}$ and $B_{\theta}$ which gives
 $\bar P=R(\eta)\Theta(x)$ equals to a constant or
\\
$$\bar P(\eta,x)\,
 =\,(a\eta)^{1/2}J_{n+1/2}(a\eta)\Theta(x)\,
 =\,C\,\,\,.\eqno(20b)$$
\\
In other words, the nested field lines are given by the contours
 of $\bar P(\eta,x)$ on the $(r-\theta)$ plane.
 
As an example, the field lines for $n=4$ and $b^2=0$ are shown in
 Fig.(1) for axisymmetric secondary plasmoids. We have taken
 $a\eta_{0}=8.2$ at the first zero of the spherical Bessel function
 where $\eta_{0}$ is the radial label of the plasma boundary.
 The horizontal axis is $x=0$ or $\theta=\pi/2$ and the vertical
 axis is $x=+1$ or $\theta=0$. For $n=4$, $\Theta(x)$ vanishes at
 $x=0,\pm(3/7)^{1/2},\pm 1$, and $R(\eta)$ vanishes at
 $a\eta=0.0,8.2$. These are locations where $B_{\theta}=0$ and
 $B_{r}=0$ respectively dividing the quadrant in two regions,
 as indicated in Fig.(1). These dividing lines are obtained by
 solving the contours with $C=0$ numerically. The radial grids are
 equally spaced in $\eta$. The meridian grids are equally spaced
 in $x$. By converting to $\theta$ through $x=\cos\theta$,
 it generates an uneven grid distribution in $\theta$ that appears
 in Fig.(1).
 Closed field lines are also shown in each region.
 Negative value contours of $C=-0.3,-0.5,-0.7$
 show the field lines in the region adjacent to the axis of $x=0$,
 and positive value contours of $C=+0.3,+0.5$ show the field lines
 in the region adjacent to the axis of $x=+1$.
 At the topological center of each region, we have $\Theta(x)$
 maximum and $R(\eta)$ maximum, thereby giving $B_{\theta}=0$
 and $B_{r}=0$ at the same location. Since
\\
$$2\pi r\sin\theta B_{\phi}\,=\,2\pi aP\,\,\,,$$
\\
the center has the maximum of line integral of $B_{\phi}$ about
 the axis of symmetry. Adjacent secondary plasmoids have oposite
 signs of $B_{\phi}$. Should we take $a\eta_{0}=11.7$ as the second
 zero of the spherical Bessel function, one additional shell of
 secondary plasmoids would be added to Fig.(1). Furthermore,
 because of the $(1-x^2)$ factor in Eq.(19b'), $\Theta(x)$ is
 null at the poles with $x=\pm 1$ where the magnetic field is
 purely radial. If $n\gg 1$, the null at the poles is tightly
 surrounded by a polar lobe. Along the polar lobe, the embedded
 belt plasmoids are tightly wounded about the polar axis, and we
 name them secondary plasmoids.

We are now ready to examine the general case of $b^2\neq 0$.
 This case of $b^2\neq 0$ is known as the finite $\beta$ case where
 $\beta$ is the ratio of plasma pressure to magnetic pressure.
 The presence of the plasma pressure with finite $b^2$ distorts the
 linear solution of $\Theta(x)$. With finite $\beta$, the series
 usually has a finite value at $x=\pm 1$ which leads to singular
 magnetic fields there because of the $(1-x^2)^{1/2}$ factor in
 the denominator. Constrained by nature to regular solutions,
 $b^2$ has to be the eigenvalues such that $\Theta(x)$ remains
 null at $x=\pm 1$. To search for these eigenvalues, we evaluate
 $\Theta(x)$, Eq.(19b), at the boundary $x=+1$ for different values
 of $b^2$ with a given $n$. The eigenvalues are shown in Fig.(2)
 as the intercepts of $\Theta(+1)=0$ at 119, 261, 425 for even
 series with $n=13$. As for $n=23$, they are 200, 422.
 The even series has $\Theta(0)\neq 0$ and $d\Theta(0)/dx=0$ and
 the odd series has $\Theta(0)=0$ and $d\Theta(0)/dx\neq 0$.
 The even eigenfunctions with $n=13$ and $b^2=119$, and with
 $n=23$ and $b^2=200$, are shown in Fig.(3). With $n=13$ and
 $b^2=119$, there are seven nodes in the inverval $(0,1)$. Should we
 take the second eigenvalue $b^2=261$, one more node would be added.
 With $n=23$ and $b^2=200$, there are twelve node in $(0,1)$.
 The next eigenvalue $b^2=422$ would add one more node as well.

The fact that the plasma pressure has to be as such that it is
 the eigenvalue of the $\Theta(x)$ equation appears to be a very
 restrictive constraint for the self-similar solutions.
 Nevertheless, we note that the plasma pressure appears in the
 $\Theta(x)$ equation where the separation constant $n(n+1)$
 is an as yet unspecified free parameter. As a result, there
 will be an adequate $n$ for almost any given plasma pressure.
 For example, with $b^2=422$, instead of being the second eigenvalue
 of $n=23$, it could be the first eigenvalue of some different $n$
 that is larger than 23. The important point is that for any plasma
 pressure $b^2$, it will coincide to one of the eigenvalues of
 some $n$. The chances are that $n$ is a large integer which,
 as we will show in the next section, leads to good jet collimation.
 
\newpage
\section{Jet Collimation and Vortex Structure}
 
One important point we should point out is that, although we have
 taken a spherically symmetric radial expansion in the plasma
 velocity, the plasma density and pressure need not be symmetric,
 and much less the magnetic fields. The plasma profiles will be
 discussed later. Here, we discuss the magnetic fields that are given
 in terms of $R(\eta)$ and $\Theta(x)$ as listed in the preceeding
 section. The maximum of $\Theta(x)$ at $x=0$ gives $B_{r}=0$ on the
 equatorial acreation disk plasma where the magnetic field has only
 a $B_{\theta}$ meridian and a $B_{\phi}$ azimuthal component.
 Off the disk, due to the nodes in Fig.(3), there are also maxima
 in space, so that the magnetic fields are structured in lobes
 other than the equatorial lobe.
 
To consider the collimation of polar jets, we examine the magnetic
 field components. For $B_{\theta}$ and $B_{\phi}$ components,
 we plot the function $\Theta(x)/(1-x^2)^{1/2}$ in Fig.(4) with the
 corresponding eigenvalue $b^2=119$ for $n=13$ according to Fig.(2).
 It peaks up off the scale to (-3.0) as $x$ approaches unity, but
 plunges to null at $x=+1$. As $n$ increases, the peak edges closer to
 $x=+1$. This is clearly shown in Fig.(4) with $b^2=200$ for $n=23$.
 The peak then goes to (+3.8) in this case. As a consequence, the peak
 rises up in amplitude but narrows down in width.
 To examine $B_{r}$, we plot the function $d\Theta(x)/dx$ in Fig.(5)
 which shows that the magnetic field is purely radial on the polar
 axis. As we have said, this radial magnetic field changes sign
 as each region of $\eta_{i}$ is crossed.
 With $n=13$ and $n=23$, the peak on the polar axis is (+71)
 and (-155) respectively.
 The magnetic energy is the quadratic quantity of this, and it peaks
 even more with respect to the off axis lobes.
 The detail structures near the polar axis
 are shown in Fig.(6) for $B_{\theta}$ and $B_{\phi}$, and in Fig.(7)
 for $B_{r}$. The magnetic field, therefore, converges to the polar
 axis as $n$ increases leading to a jet structure.
 Comparing the equatorial lobe of the magnetic field with the polar
 lobe, it is clear that the polar lobe is much narrower than the
 equatorial one. This lobe pattern is similar to directional antenna
 arrays. Due to the oscillating nature of Eq.(19a), there
 are spherical toroids, or secondary plasmoids, formed by closed
 magnetic field lines along this narrow peak. The number of
 secondary plasmoids embedded in this peak measures the number
 of zeros of Eq.(19a) contained within $a\eta_{0}$. The magnitude
 of $a\eta_{0}$ also indicates the large amplitude of $B_{\phi}$
 through $\bar Q=a\bar P$ of Eq.(18a) which is itself consistent
 to the magnetic field convergence onto the polar axis.
 
To understand the magnetic structure of the jet, we plot the contours
 of $\bar P$ given by Eq.(20b) which amount to closed field lines of
 embedded secondary plasmoids. A contour plot in Fig.(8), similar to
 Fig.(1) but with $n=13$ and $b^2=119$, shows plasmoids along the polar
 cone that goes from $x=0.96$ to $x=1$, as in Fig.(7), which corresponds
 to a cone angle of about 14 degrees. The radial label goes from
 $a\eta=0$ to $a\eta=40$, and this interval is divided into seven
 regions set by the zeros of the spherical Bessel function. The first
 zero is at $a\eta=18.40$ and the subsequent zeros are approximately
 equally spaced, as indicated in Fig.(8). Negative contours of $C=-0.010$
 are plotted in the first, third, fifth, and seventh regions only.
 Likewise, positive contours of $C=+0.005$ are plotted in second,
 fourth, and sixth regions. To avoid over crowdedness, the $C=-0.005$
 contours are not plotted in the odd numbered regions, and the
 $C=+0.010$ ones are not shown in the even numbered regions.
 Once more, we recall that the magnetic field lines of these secondary
 plasmoids circle in alternating sequence. The boundary of the jet
 is given by the root of $\Theta(x)$ with $x_{j}=0.96$. At this boundary,
 the magnetic field is purely radial but alternates in sign. The other
 boundary is at $x_{j}=1.0$ where the magnetic field is also radial
 and alternates in sign opposite to that of $x_{j}=0.96$. Together with
 the roots of $R(\eta)$ at $\eta_{i}$ where the magnetic field is
 meridian, the plasmoids are magnetic vortices bounded by closed field
 lines at the border of each region. By axisymmetry, these poloidal
 magnetic contours rotate about the polar axis to generate tightly
 wounded toroids. Furthermore by axisymmetry, the toroidal azimuthal
 field lines are decoupled from the bipolar poloidal field lines.
 The field lines are, therefore, two-dimensional.
 As in the discussions concerning Fig.(1), this amounts to the $m=0$
 mode in the Fourier transform $e^{im\phi}$ of the toroidal dependence.

The magnetic structure of this spherical temporal self-similar model,
 which shows a sequence of magnetic toroids along the jet, differs
 from the continuous helical field line structure in the cylindrical
 spatial self-similar model, complemented by numerical simulations
 [Casse 2004]. To understand the differences, the spatial cylindrical
 model has an imposed initial magnetic configuration everywhere in space.
 Jets are formed by transporting the disk plasma from the disk plane
 to the axis through magnetocentrifugal action. They are maintained
 along the axis by bringing disk plasma continuously in steady
 state to overcome transport losses. In this description, spatially
 self-similar jets in $z/r$ were formed in their present position
 in the distant past, and are maintained there continuously at the
 present and in the future.
 
As for the temporal self-similar spherical model, the disk plasma
 is first accreted to the galactic nucleus which builds up the plasma
 pressure there. While the plasma is in this bounded region,
 self-organization is nourished through dissipations to self-similar
 structures in $\eta=r(t)/y(t)$ calculated here, including the jets.
 The force-free configuration and the vortex nature of the magnetic
 field are akin to the quadratic quantities of magnetic helicity and
 cross helicity of the ideal MHD system. We note that $\eta$ is time
 invariant, such that the temporal self-similar $\eta$ profiles in
 magnetic field and plasma parameters maintain their forms at the very
 beginning. Later on, the self-similar MHD plasma errupts from the
 center outward, and this entire structure expands into the previously
 void space as time progresses, and continues to do so until the plasma
 density and magnetic field intensity get so low that they fade into
 space light years away. The bounded state and the erruption are
 described by the time evolution function $y(t)$, and it will be solved
 consistently in the next section.

\newpage
\section{Evolution Function}

The $r$ component of the momentum equation reads
\\
$$\rho[{\partial v\over\partial t}
 +v{\partial v\over\partial r}]
 +{d\over dr}p(r,P(r,\theta))
 +\rho{GM\over r^2}\,$$
 
$$=\,-{1\over\mu}({1\over r\sin\theta})^2
 {\partial P\over\partial r}
 \{{\partial^2 P\over\partial r^2}
 +{1\over r^2}\sin\theta{\partial\over\partial\theta}
 ({1\over\sin\theta}{\partial P\over\partial\theta})
 +Q{\partial Q\over\partial P}\}\,\,\,.\eqno(21)$$
\\
The term $dp/dr$ on the left side refers to radial derivatives
 for the explicit dependence and the implicit dependence in $P$.
 Making use of the $\theta$ component, Eq.(15a), the equation
 above becomes
\\
$$\rho[{\partial v\over\partial t}
 +v{\partial v\over\partial r}]
 +{d\over dr}p(r,P(r,\theta))
 +\rho{GM\over r^2}\,$$
 
$$=\,-{1\over\mu}({1\over r\sin\theta})^2
 {\partial P\over\partial r}
 \{-\mu r^2\sin^2\theta
 {\partial p\over\partial P}\}\,
 =\,+{\partial P\over\partial r}{\partial p\over\partial P}
 \,\,\,.$$
\\
The right side is just the radial derivative of plasma pressure
 on the implicit dependence in $P$. This cancels the corresponding
 term on the left side leaving only the explicit radial derivative
\\
$$\rho[{\partial v\over\partial t}
 +v{\partial v\over\partial r}]
 +{\partial p(r,P)\over\partial r}
 +\rho{GM\over r^2}\,=\,0\,\,\,.$$
\\
In terms of self-similar parameters, this equation reads
\\
$$y^2{d^2y\over dt^2}
 +{1\over\eta}
 \{y^{4-3\gamma}{1\over\bar\rho}
 {\partial\bar p\over\partial\eta}
 +{GM\over\eta^2}\}\,
 =\,y^2{d^2y\over dt^2}
 +{1\over\eta}
 \{{1\over\bar\rho}
 {\partial\bar p\over\partial\eta}
 +{GM\over\eta^2}\}\,
 =\,0\,\,\,.\eqno(22)$$
\\
With $\alpha$ as the separation constant, and $H$ as an integration
 constant, the evolution function is, therefore, described by
\\
$$y^2{d^2y\over dt^2}\,=\,+\alpha\,\,\,,$$

$$[{1\over 2}({dy\over dt})^2+{\alpha\over y}]\,
 =\,H\,\,\,,\eqno(23a)$$

$${dy\over dt}\,
 =\,\pm[2(H-{\alpha\over y})]^{1/2}\,\,\,.\eqno(23b)$$
\\

To understand the meaning of $\alpha$, we note that plasma
 acceleration in Lagrangian coordinate is
\\
$${dv\over dt}\,=\,\eta {d^2y\over dt^2}\,
 =\,{\alpha\eta\over y^2}\,=\,{\alpha\eta^3\over r^2}\,\,\,.$$
\\
A negative $\alpha$ means an outward decelerating flow, or an inward
 accelerating flow. The deceleration gets smaller as $y$ or as $r$
 gets larger, the acceleration gets larger as $y$ or as $r$ gets
 smaller. As for the meaning of $H$, we take the limit $y=\infty$ which
 gives $H/4\pi=(v^2/2)/4\pi\eta^2$ by using $r=\eta y$. Physically,
 this is the asymptotic radial kinetic energy per unit mass per unit
 area of a Lagrangian fluid element. For circular orbits, we would
 have $H=0$. Furthermore, we can rewrite Eq.(23a) making use the above
 physical expression of $\alpha$ to get
\\
$$\rho\eta^{2}H\,
 =\,{1\over 2}\rho v^{2}+(\hat r\cdot\vec F)r\,\,\,,$$
\\
where $\vec F$ denotes all the forces on the right side of Eq.(2).
 It is clear that $H$ measures the total energy of the fluid element.
  
To consider the accretion disk magnetospheric plasma, we take
 $\alpha=-|\alpha|<0$ negative for an inward axcelerating flow.
 With $H=0$ as the boundary condition at infinity, we get
\\
$${dy\over dt}\,=\,\pm({2|\alpha|\over y})^{1/2}\,
 =\,-({2|\alpha|\over y})^{1/2}\,\,\,.\eqno(24a)$$
\\
The two signs for the square root correspond to the outward and inward
 flows. Taking the negative sign for the inward flow, the equatorial
 accretion disk plasma starts at infinity with an asymptotically
 zero radial velocity and ends near the center with a large influx.

As to describe the polar jets, we again take $\alpha=-|\alpha|<0$
 negative. We consider $H=\pm|H|$ which lead to
\\
$${dy\over dt}\,
 =\,\pm[2(\pm|H|+{|\alpha|\over y})]^{1/2}\,\,\,.\eqno(24b)$$
\\
For $H=+|H|>0$ and taking the positive sign on the square root,
 this gives a decelerated outward flow.
 The deceleration gets smaller as $y$ or as $r$ gets larger,
 with the terminal velocity given by $v=\eta dy/dt=\eta (2H)^{1/2}$.
 For $H=-|H|<0$, the outward flow would stop at $y=|\alpha|/|H|$
 where $dy/dt=0$. This $H=-|H|<0$ outward flow would be followed by
 an accelerated inward flow, should we take the negative sign on
 the square root, so that the system would be periodic and bounded.
 We, therefore, see that the system would make a transition from a
 bounded to an unbounded state when $H$ goes from negative to positive.
 The bounded state allows us to define the boundaries of the
 radial label $(0,\eta_{0})$ such that $(0<\eta<\eta_{0})$.
 The solutions of the evolution function in Eq.(24b) suggest that
 jets are a result of an erruptive process. This process is fed by
 plasma accretion from the galactic disk. As pressure builds up at
 the center, the plasma begins to oscillate, or to pulse, as described
 by the bounded solution with $H=-|H|<0$. We believe that it is
 during this phase that self-organized and self-similar structures
 in plasma parameters and magnetic fields would be nourished through
 dissipations. The dynamics is given by self-similar solutions
 where the magnetic fields are self-consistent to the plasma
 pressure that plays the role of the eigenvalue. The jet structure
 emerges along the polar axis as the eigenfunction of the magnetic
 field. By further pressure built-up and energy influx, $H=+|H|>0$
 becomes positive, and the structure eventually errupts.

\newpage
\section{Rayleigh-Benard Cells}

Now we have presented the general structure of galactic jets.
 It begins with an influx of plasmas from the accretion disk
 which drives up the plasma pressure at the galactic nucleus.
 The MHD plasma responses to self-organization by oscillating,
 or pulsating, periodically in time. As $H$ goes from negative
 to positive, the periodic mode goes to an erruptive mode.
 The result of self-organization is a structured configuration
 in space. This structure is not unique. For galactic jets, we
 have solved the system with the method of separation of variables
 as such that it ressembles to observations. This structure contains
 basically convective cells in space where magnetic field lines
 in adjacent cells rotate in opposite sense. Although such a
 structured configuration with consistent evolution function is
 in accordance with the time-dependent MHD equations, there is no
 mention of the initial configuration that could lead to such
 self-similar states. Consequently, such a highly organized complex
 structure could be regarded as an artificial result due to special
 mathematical constructions, which might not have any relevance to
 the physical jet system.
 In order to bring more reality to this analytic result, we recall
 the classical case of Rayleigh-Benard fluid self-organization.
 Consider a thin layer of oil heated from below, observations tell
 us that this simple homogeneous configuration will develope an array
 of identical hexagonal convective cells if the temperature gradient
 across the layer is sufficiently large. The velocity streamlines
 of adjacent cells rotate in opposite sense. Should the initial state
 of this oil layer be altered by arbitrary fluctuations, same
 convective cells would still appear after the initial fluctuations
 are dissipated. Numerically, such self-organized complex structure
 could be reached from the simple homogeneous initial configuration
 if the code is adequately pushed in the correct direction.
 
Guided by this fluid example, the self-similar jet structure could
 be regarded as the Rayleigh-Bernard equivalent in the MHD system.
 We could choose an initial state by taking $n=1$ in our solutions
 with plasma bounded between an outer sphere $\eta_{0}$ and an inner
 sphere $\eta_{*}$. This would give a large global long wavelength
 structure in the spherical layer. We consider an equilibrium state
 with $\alpha=-|\alpha|<0$ and $H=-|H|<0$ with $y=|\alpha|/|H|$ and
 $dy/dt=0$. Let us take this moment as $t=0$.
 An energy source in terms of pressure is supplied by the accretion
 disk to pump the MHD plasma at the lower boundary $\eta_{*}$.
 At some moment, $y$ begins to depart from its equilibrium to fall
 towards the center due to perturbations. Like in the oil layer case,
 this could trigger convections so that the long wavelength global
 structure cascades to short wavelength structures accepted by the
 system. Cenvective cell scaling becomes smaller as $n$ gets larger.
 Different from the flat layer fluid case, the convective cells in
 this spherical MHD layer are not identical, since there is a focusing
 effect to the polar axis. 

\newpage
\section{Mass Density Profile}

As for the spatial part, we have
\\
$${1\over\bar\rho}
 {\partial\bar p\over\partial\eta}
 +{GM\over\eta^2}\,=\,-\alpha\eta\,\,\,.$$
\\
With $\bar p$ given by Eq.(18b), the above equation gives
\\
$$\bar\rho(\eta,\bar P)\,
 =\,(\eta^{-3})({4\bar p_{2}(\bar P)\over\alpha\eta^3+GM})\,
 =\,(\eta^{-3})({4\bar p_{2}(\bar P)\over GM})\,$$
 
$$=\,(\eta^{-3})({4\over GM})(b'^2\bar P^2+\bar C)\,
 =\,\bar\rho_{1}(\eta)\bar\rho_{2}(\bar P)\,\,\,.\eqno(25)$$
\\
In our present case, we are considering a spherical shell domain
 that the radial label $\eta=0$ is excluded. The negative powers
 of $\eta$ in plasma pressure and mass density do not cause
 singularity. We have set $\alpha=0$, with vanishing net force
 on the flow, to obtain the second equality in Eq.(25) to be
 compatible with the functional form of $F=F(P)$ with $\gamma=4/3$.
 Nevertheless, we could relax this condition to $\alpha\approx 0$
 such that $\alpha\eta^3<<GM$. To understand the implication of
 such inequality, we remark that
\\
$${\alpha\eta^3\over GM}\,=\,{dv/dt\over GM/r^2}\,<<\,1
 \,\,\,.\eqno(26)$$
\\
Therefore, it is just the acceleration of the flow to that of
 the central mass gravitational acceleration. Since $dv/dt$ gets
 smaller as $r$ gets larger with $1/r^2$ scaling, this condition
 establishes that self-similar configurations can be organized,
 and bounded oscillations of the $H=-|H|<0$ case can take place,
 as long as the plasma acceleration is much less that the gravitational
 acceleration. To close the entire self-similar system, we now come
 to the functional $F(P)$ for the adiabatic equation of state.
 With the results in Eq.(18b) and Eq.(25) using $\alpha\approx 0$,
 we have
\\
$$\bar F(\bar P)\,
 =\,{\bar p_{2}(\bar P)\over\bar\rho_{2}^{\gamma}(\bar P)}\,
 =\,({GM\over 4})^{4/3}\bar p_{2}^{-1/3}(\bar P)\,
 =\,({GM\over 4})^{4/3}(b'^2\bar P^2+\bar C)^{-1/3}
 \,\,\,.\eqno(27)$$
\\

To understand the plasma structure of the jet, we note that the
 plasma pressure and mass density are $b'^2\bar P^2(\eta,x)+\bar C$
 dependent which means positive definite $b'^2C^2+\bar C$ dependent,
 according to Eq.(18b) and Eq.(25) respectively.
 The plasma pressure and the mass density have their minimum at
 $\bar P(\eta,x)=(a\eta)^{1/2}J_{n+1/2}(a\eta)\Theta(x)=0$.
 This minimum is positive nonzero because of the positive integration
 constant $\bar C$ in Eq.(18b). To discuss the plasma structure in
 the jets, we reproduce the part of Fig.(3) in the range $0.94<x<1.0$
 in Fig.(9). The segmented line for $n=13$ has $\Theta(x)=0$ at the
 cone boundary $x=0.96$ and at the cone center $x=1.0$. In between,
 there is a minimum at $x=0.985$. Because of the quadratic dependence
 $\Theta^{2}(x)$, these features correspond to a cavity-like structure
 for the plasma pressure and mass density of the jet, compatible
 to numerical simulations of the time-dependent dissipative MHD
 equations [Casse 2004, Zanni et. al. 2004]. Other than the zeros of
 $\Theta(x)$, $\bar P(\eta,x)$ also vanishes at radial locations where
 $J_{n+1/2}(a\eta)=0$. With the quadratic dependence
\\
$$R^{2}(\eta)\,=\,(a\eta)J^{2}_{n+1/2}(a\eta)\,\,\,,$$
\\
there are ripples for plasma pressure and mass density along the
 radial label. The radial function has two contributions. The part
 of the Bessel function $J_{n+1/2}(a\eta)$ has an oscillating nature
 with decreasing amplitude, while the part of $(a\eta)$ has an
 increasing amplitude that helps to maintain the ripple level of
 $R^{2}(\eta)$. The peaks of these ripples are at the topological
 center of each plasmoid which is the magnetic axis of the toroid
 where $B_{\phi}$ has a maximum line integral. These ripples are
 also seen in numerical simulations [Zanni et. al. 2004].
 It is quite surprising that the results of this spherical temporal
 self-similar MHD model agree rather well qualitatively with numerical
 simulations of the time-dependent MHD equations for the cylindrical
 magnetocentrifugaling model. Both of them give cavity structures in
 the transverse direction and ripple structures in the longitudinal
 direction.

The polar jets, therefore, have a periodic and approximately equally
 spaced concentration of plasmoids in its longitudinal direction,
 except the first region which is more extensive. The closed magnetic
 field lines for adjacent plasmoids rotate in opposite sense. The
 plasma pressure and mass density have a hollow conic structure
 with ripples along the radial direction. Such periodic structure
 happens to be compatible with radio observations where high intensity
 islands are mapped. These islands are usually thought to be periodic
 ejections of mass from the accretion disk. According to our model,
 they are rather the internal spatial arrangments of an expanding jet
 driven by an erruptive event.

\newpage
\section{Discussions and Conclusions}

One of the main objections of self-organized plasmoid representation
 in free space in the absence of an adequate boundary is that it
 apparently violates the Virial theorem which states that
 [Schmidt 1966]
\\
$${1\over 2}{d^2 I\over dt^2}
 +\int x_{k}{\partial G_{k}\over\partial t}dV\,
 =\,2(T+U)+W^{E}+W^{M}
 -\int x_{k}(P_{ik}+T_{ik})dS\,\,\,,\eqno(28)$$
\\
where $I$ is the moment of inertia of the plasmoid, $G$ is the
 momentum density of the electromagnetic field, $T$ and $U$ are the
 kinetic and thermal energies of the plasma, $W^{E}$ and $W^{M}$
 are the electric and magnetic energies in the volume, $P_{ik}$
 and $T_{ik}$ are the plasma and electromagnetic stress tensors.
 Taking the volume to cover the entire plasma and field, the surface
 term on the right side vanishes. In laboratory plasmas, this surface
 could be the machine vessel. Should the plasmoid be stationary,
 the volume term on the left side would be null, and the moment of
 inertia would be accelerating since the terms on the right side are
 all positive definite. While this statement has no conflict with
 the unbounded solutions, it apparently contradicts the stationary
 state of the bounded solutions. Nevertheless, this argument has
 overlooked the asymptotically bounded nature of the $H=-|H|<0$
 plasmoid state. In this asymptotic case, we have
 $dI/dt=(dI/dy)(dy/dt)=0$ so that $I$ is stationary. The fact that
 $d^2I/dt^2>0$ implies that $I$ is at an asymptotic minimum, not an
 acceleration of $I$, which complies with the Virial theorem.

The classical accretion-ejection model of Blandford and Payne [1982]
 is a time-independent stationary state model with spatial self-similar
 MHD solutions in cylindrical geometry $(r,\phi,z)$ with Alfv\'{e}nic
 plasma flow velocity plus a rotation, all with Keplerian scaling.
 In this model, the jets are formed by convecting the magnetospheric
 disk plasmas from the disk plane to the axis by magnetocentrifugal
 action through the magnetic field lines with low inclination angles
 to the disk plane. The angular momentum of the plasma on the disk
 plane is focused to the axis. The jets were put in place in the
 distant past according to the spatial self-similar solutions in $z/r$,
 and are maintained there by continuously transporting disk plasmas
 to the axis to sustain the axial outflow. Collimation to the axis
 is accomplished by magnetic hoop force.

Here, we have taken a dynamic view where jets are the consequences
 of erruptive events, based on time-dependent MHD equations in
 spherical geometry $(r,\theta,\phi)$. The disk plasmas are accreted
 to the galactic nucleus where plasma pressure is built up to cause
 an erruption. The radially symmetric expanding velocity interacts
 the plasma with the magnetic field self-consistently through the MHD
 equations to generate spatial structures. Due to the existence of
 multiple quadratic invariants in the absence of dissipations, MHD
 systems have the tendency of developing self-organized and self-similar
 states through dissipative processes. The force-free configuration
 and the vortex nature of the magnetic field are akin to the quadratic
 invariants of the magnetic helicity and cross helicity of the MHD
 system. For these reasons, although temporally self-similar solutions
 are only a subset of general time-dependent MHD solutions, these
 self-similar configurations are prone to develope in natural phenomena.

We, therefore, describe these spatial structures by self-similar
 temporal solutions in $\eta=r/y$. In this self-similar spherical model,
 consistent self-similar representations of plasma pressure, mass
 density, and magnetic fields are worked out for the $\gamma=4/3$
 BC Low model, with special emphasis on the finite $\beta$ case.
 The spatial distribution of the magnetic field is described by
 an equation where the plasma pressure acts as the eigenvalue.
 The nature of this eigenvalue equation is as such that the magnetic
 field gets converged to the polar axis as a response to the high
 plasma pressure directly related to the eigenvalue.
 Since the separation constant $n(n+1)$ in this eigenvalue equation
 is a free parameter that as yet to be specified, there will be
 an adequate $n$ for almost any given plasma pressure.
 Although the radial plasma velocity is isotropic, the spatial
 structures are not. They could be highly collimated along
 the axial direction, and expand into the previously void space
 as time progresses.

Although other types of solutions are permitted, we have specifically
 examined the solutions that bear resemblance with jet features with
 $n\gg 1$ and $a\eta_{0}\gg 1$. Since plasma and magnetic field are
 frozen into each other, plasma outflow is also collimated to the polar
 axis. The existence of regions of closed field lines along the primary
 polar magnetic lobe permits secondary plasmoids be embedded in it.
 These secondary plasmoids appear to be compatible to the observed
 islands of radio intensities. The time evolution function of the
 radial velocity consistent to the temporal self-similar solutions
 has different types of solutions according to the sign of $H$.
 Although the equatorial disk magnetospheric plasma is not addressed
 here, the accelerated accretion of plasma influx could be modelled
 with $H=0$ as the boundary condition at infinity. As for the polar
 jets, $H<0$ gives a bounded oscillating, or pulsating, solution.
 This bounded stage, we believe, nourishes the self-organized and
 self-similar states.
 With $H>0$, it gives an unbounded expanding solution with a high
 terminal velocity. It is apparent that plasma pressure due to
 accretion is the prime driving force that determines the value of $H$.
 The bounded oscillation would make a transition to the unbounded
 expansion as $H$ goes from negative to positive. According to our model,
 jet structures are, therefore, considered as a spatial configuration
 that has been expanding continuously into space.

\clearpage
\begin{figure}
\plotone{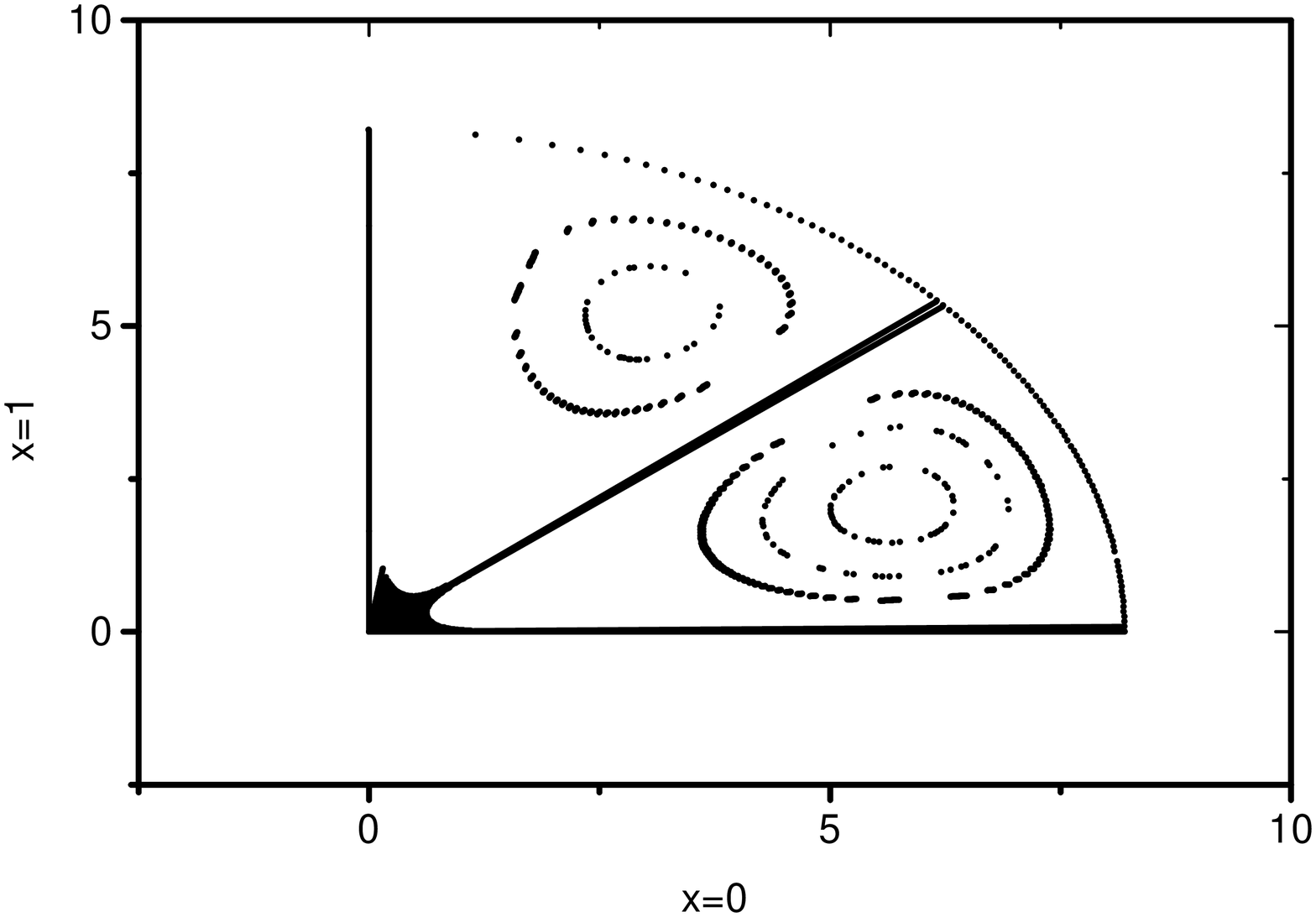}
\caption{The field lines on the $(r-\theta)$ plane are plotted in
 one quadrant with $a\eta_{0}=8.2$ as the first zero of the spherical
 Bessel function. The quadrant is divided into two regions by $C=0$
 with plasmoids having $C=-0.3,-0.5,-0.7$ and $C=+0.3,+0.5$.}
\label{fig.1}
\end{figure}

\clearpage
\begin{figure}
\plotone{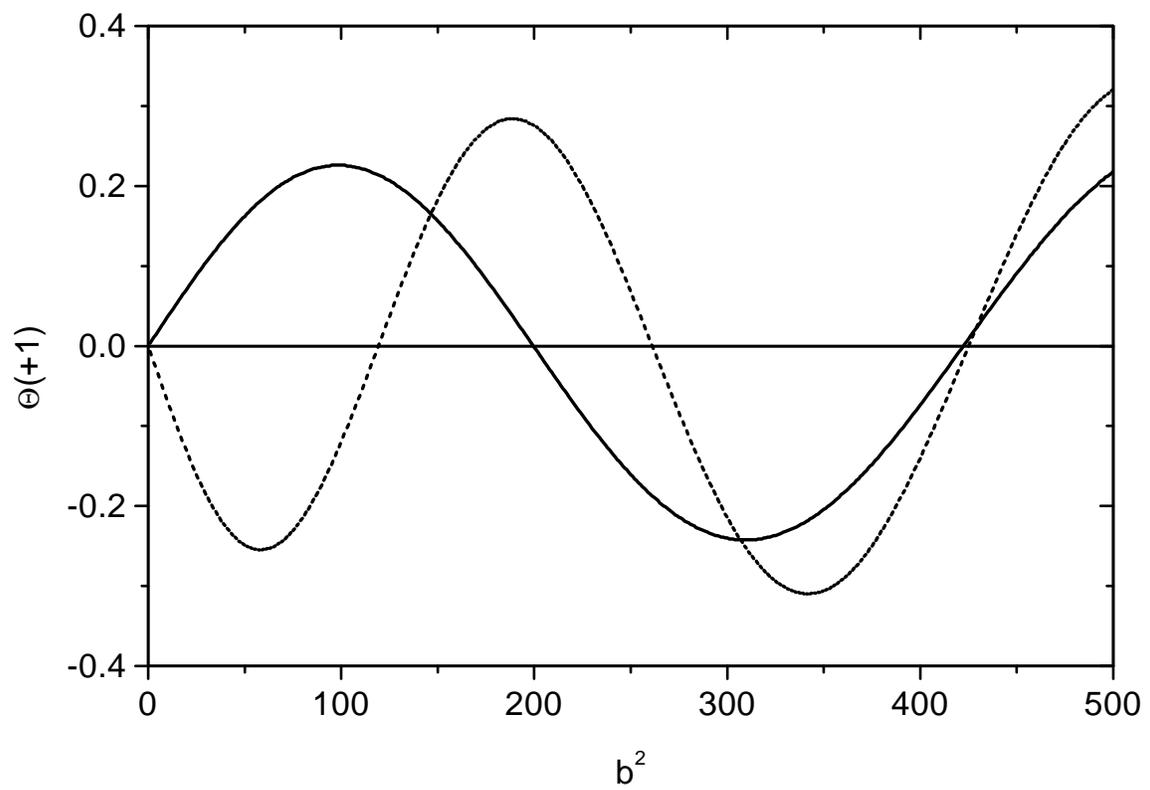}
\caption{The function $\Theta(x)$ at $x=+1$ is plotted against
 the parameter $b^2$ to identify the eigenvalues for $n=13$
 in segmented line and $n=23$ in solid line.}
\label{fig.2}
\end{figure}

\clearpage
\begin{figure}
\plotone{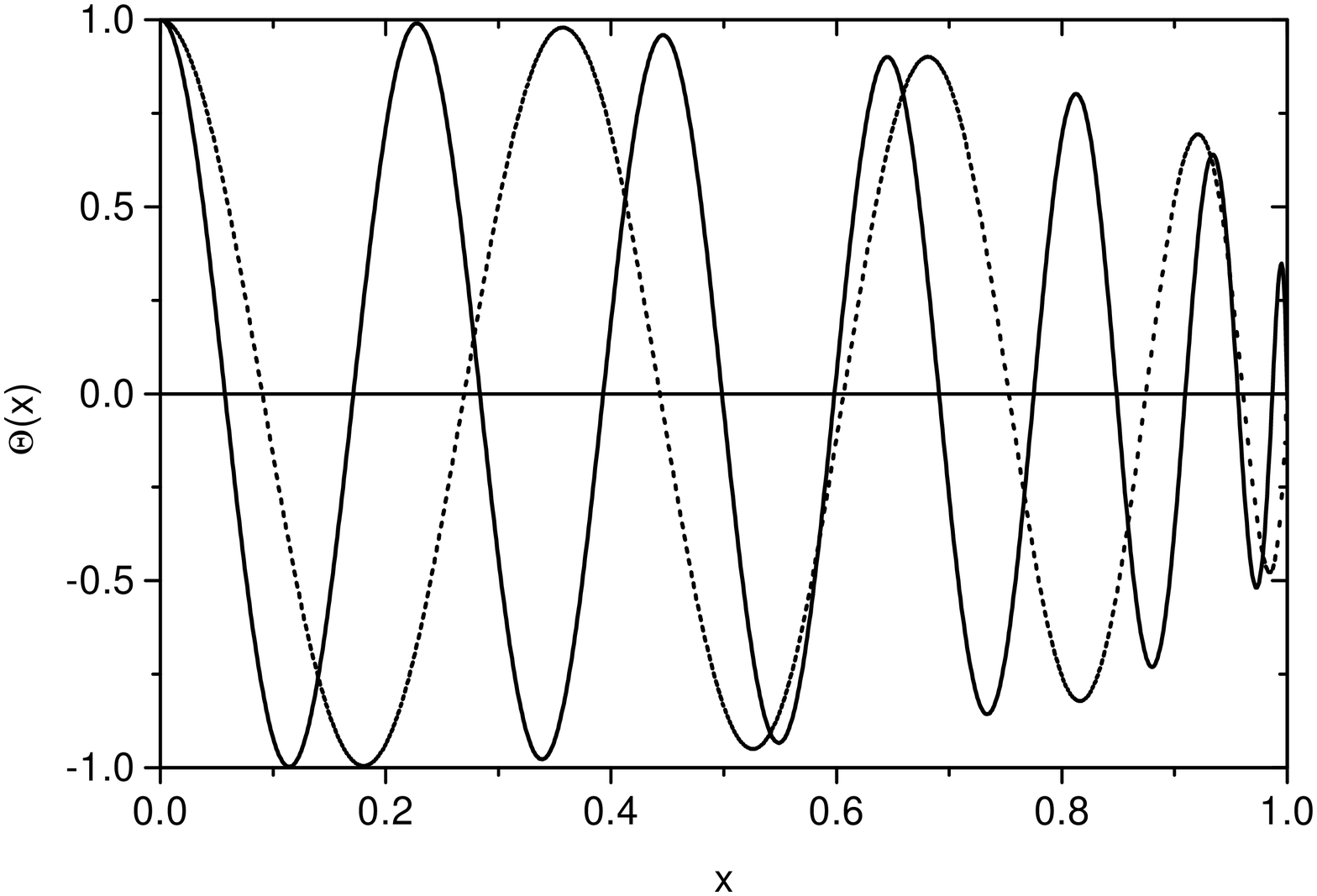}
\caption{The eigenfunction $\Theta(x)$ is plotted
 against $x$ with eigenvalue $b^2=119$ for $n=13$ in segmented
 line and $b^2=200$ for $n=23$ in solid line to show
 its dependence on $n$.}
\label{fig.3}
\end{figure}

\clearpage
\begin{figure}
\plotone{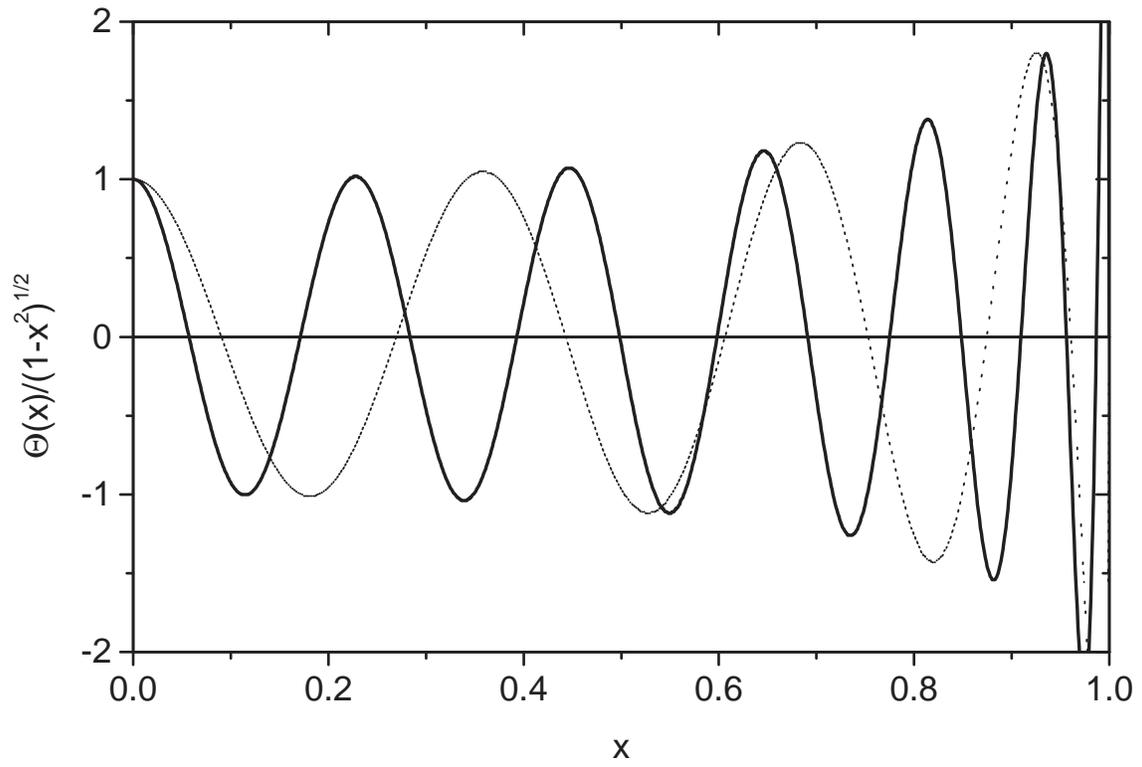}
\caption{The function $\Theta(x)/(1-x^2)^{1/2}$ is plotted
 against $x$ with eigenvalue $b^2=119$ for $n=13$ in segmented
 line and $b^2=200$ for $n=23$ in solid line to show the structure
 of $B_{\theta}$ and $B_{\phi}$ in space.}
\label{fig.4}
\end{figure}

\clearpage
\begin{figure}
\plotone{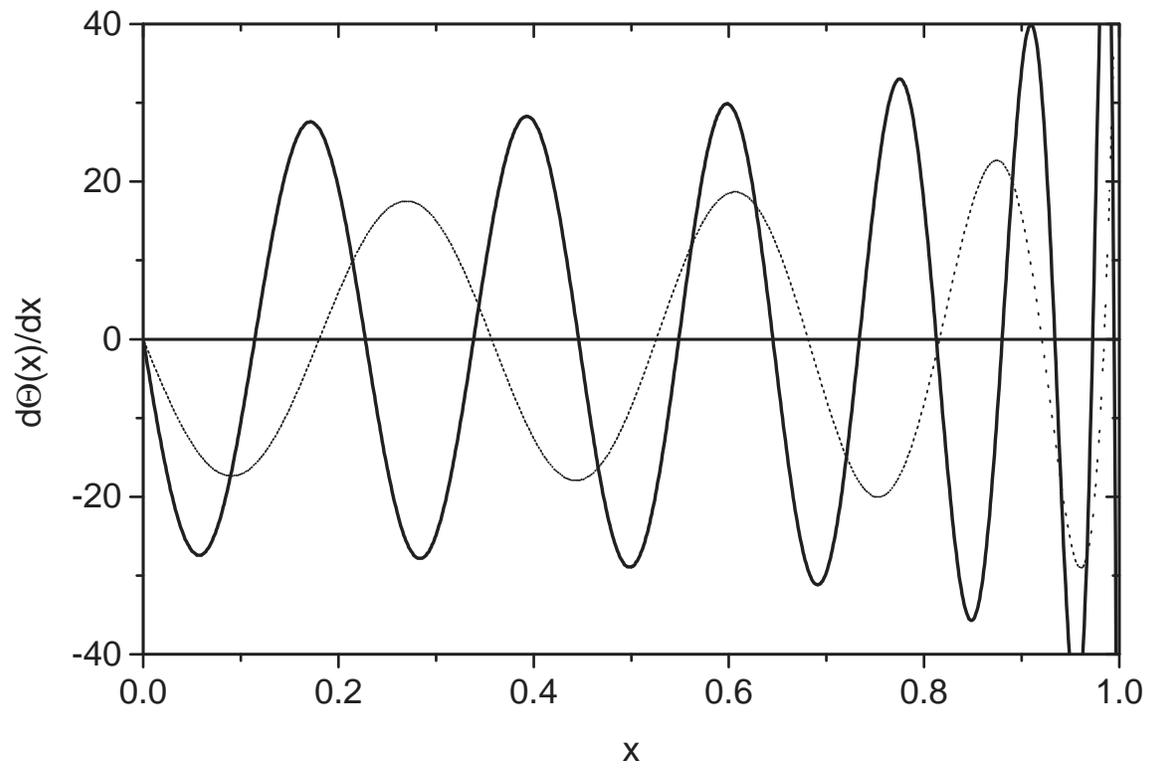}
\caption{The function $d\Theta(x)/dx$ is plotted
 against $x$ with eigenvalue $b^2=119$ for $n=13$ in segmented
 line and $b^2=200$ for $n=23$ in solid line to show the structure
 of $B_{r}$ in space.}
\label{fig.5}
\end{figure}

\clearpage
\begin{figure}
\plotone{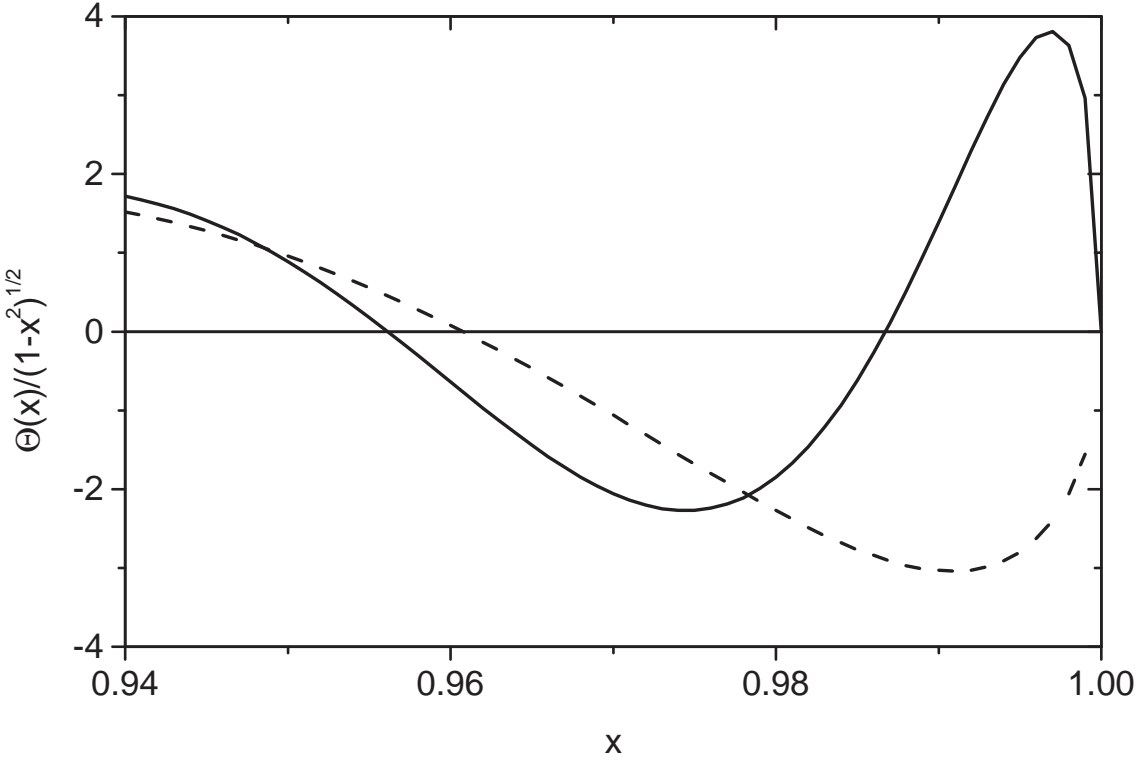}
\caption{The function $\Theta(x)/(1-x^2)^{1/2}$ is plotted
 against $0.94<x<1.0$ with eigenvalue $b^2=119$ for $n=13$ in segmented
 line and $b^2=200$ for $n=23$ in solid line to show the detail
 structure of $B_{\theta}$ and $B_{\phi}$ near the polar axis.}
\label{fig.6}
\end{figure}
 
\clearpage
\begin{figure}
\plotone{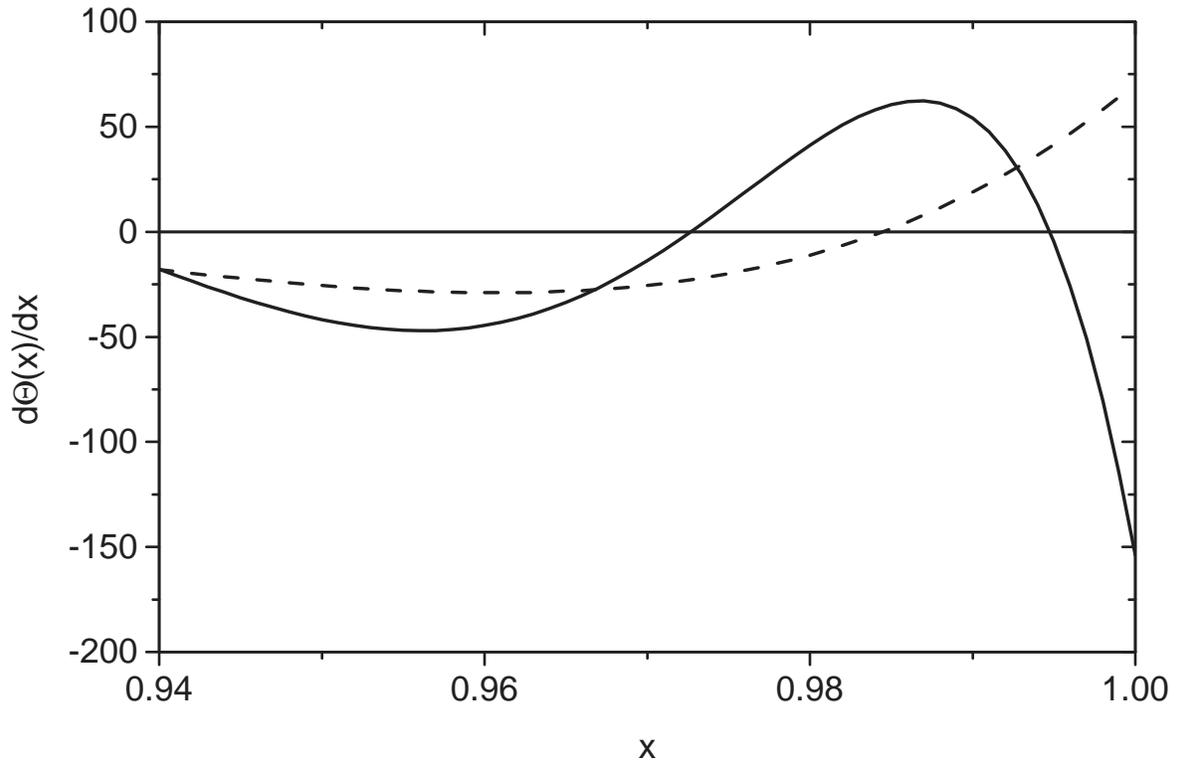}
\caption{The function $d\Theta(x)/dx$ is plotted
 against $0.94<x<1.0$ with eigenvalue $b^2=119$ for $n=13$ in segmented
 line and $b^2=200$ for $n=23$ in solid line to show the detail
 structure of $B_{r}$ near the polar axis.}
\label{fig.7}
\end{figure}

\clearpage
\begin{figure}
\plotone{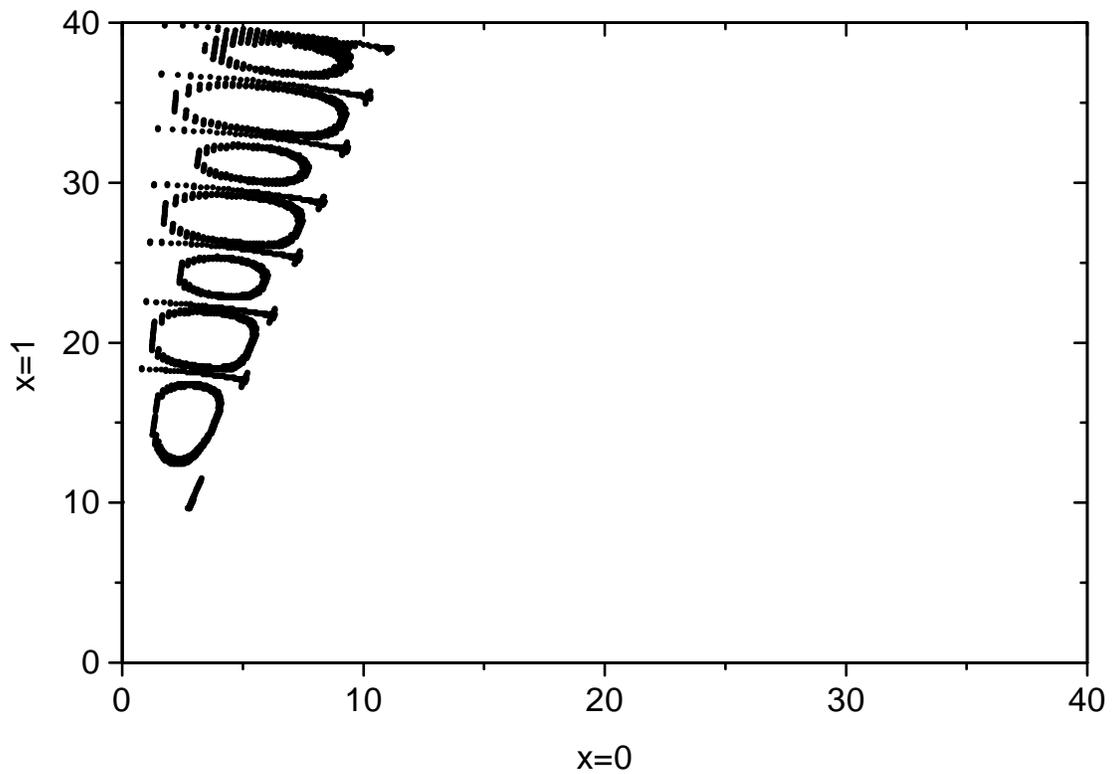}
\caption{The field lines on the $(r-\theta)$ cone around the polar
 axis are plotted with $a\eta_{0}=40$ and a not so large $n=13$
 for better resolution. The regions are divided by $C=0$ with
 plasmoids having $C=-0.010$ and $C=+0.005$.}
\label{fig.8}
\end{figure}

\clearpage
\begin{figure}
\plotone{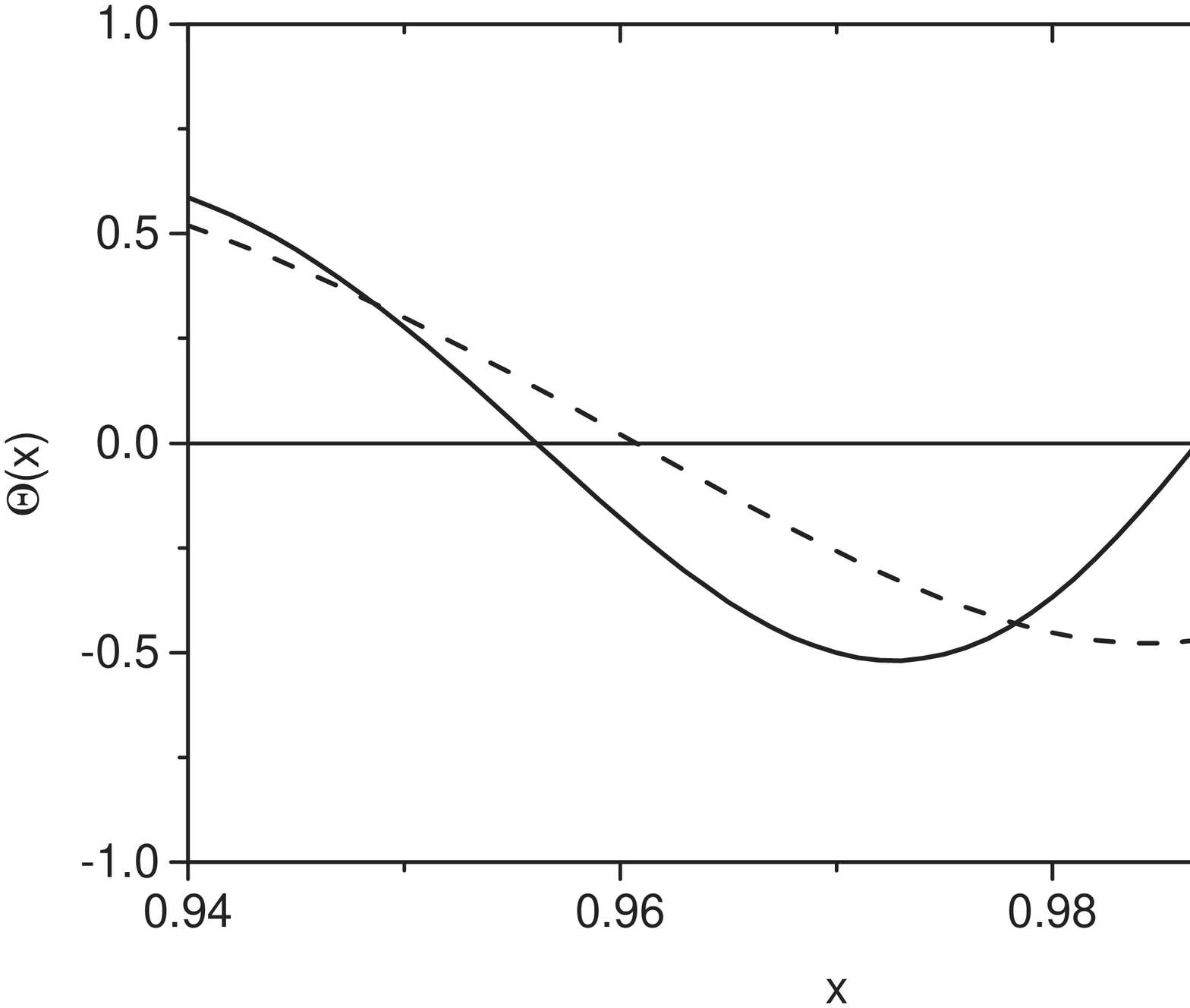}
\caption{The eigenfunction $\Theta(x)$ is plotted
 against $0.94<x<1.0$ with eigenvalue $b^2=119$ for $n=13$ in segmented
 line and $b^2=200$ for $n=23$ in solid line to show
 its dependence on $n$.}
\label{fig.9}
\end{figure}

\end{document}